\documentclass[amsmath,amssymb,prb,twocolumn]{revtex4}

\usepackage{graphicx}
\usepackage{amsmath,amssymb,amsfonts}
\usepackage[latin1]{inputenc}

\begin{document}

\title{Strong polarization-induced reduction of addition
       energies in single-molecule nanojunctions}
\author{Kristen Kaasbjerg and Karsten Flensberg}
\affiliation{Nano-Science Center, Niels Bohr Institute, University of Copenhagen,
DK-2100 Copenhagen, Denmark}

\begin{abstract}
We address polarization-induced renormalization of molecular
levels in solid-state based single-molecule transistors and
focus on an organic conjugate molecule where a surprisingly
large reduction of the addition energy has been observed. We have
developed a scheme that combines a self-consistent solution of a
quantum chemical calculation with a realistic description of the
screening environment. Our results indeed show a large reduction, and
we explain this to be a consequence of both (a) a reduction of the
electrostatic molecular charging energy and (b) polarization induced
level shifts of the HOMO and LUMO levels.
Finally, we calculate the charge stability diagram and explain at a
qualitative level general features observed experimentally.
\end{abstract}

\maketitle

\begin{figure}[!b]
  \centering
  \includegraphics[scale=1.]{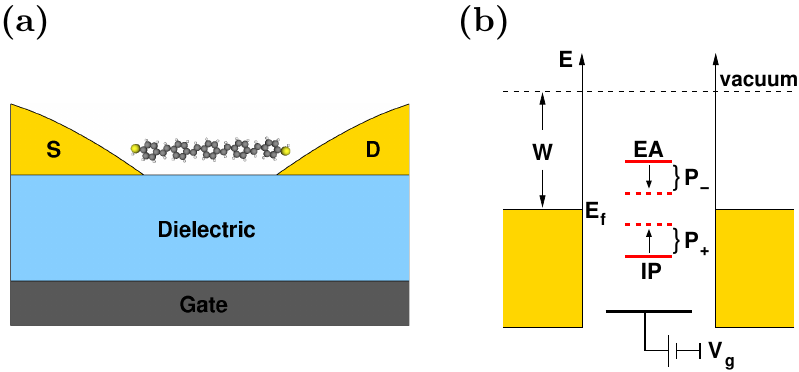}
  \caption{\label{fig:fig1}
    (a) Schematic illustration of a three-terminal single-molecule transistor.
    (b) Energy level alignment in the molecular junction showing the
    position of the molecular ionization potential (IP) and electron
    affinity (EA) levels with respect to the work functions of the metallic
    leads. Also the gate electrode with voltage $V_g$ which couples to
    EA and IP is schematically shown.
    The polarization shifts $P_+$ and $P_-$ of the levels
    due to the junction environment are indicated.}
\end{figure}

The recent experimental progress in single-molecule electronics
has resulted in the realization of the three-terminal molecular
single-electron transistor
(SET)~\cite{Liang-Nature,Park-Nature,Bjornholm-OPV5,Natelson-Kondo,
vanderZant-Vibration,vanderZantPaaske,Yao-TrimetalMolecule,Bjornholm-OPV3,
Natelson-NovelMaterials,Zant-PhysStatSol}
shown schematically in Figure~\ref{fig:fig1}a. The experimental
realizations have been based on a variety of techniques
including junctions made by electromigration, mechanical break
junctions and cryogenic nanogap fabrication. Many indications of the molecule being
part of the active transport pathway through the junction has been observed.
An example is the observation of the molecular vibrational excitations,
which serve as a fingerprint for the molecule~\cite{vanderZant-Vibration}.
There remain, however, several unresolved issues in single-molecule
transport both in the strong coupling limit, where coherent
transport theories seems to strongly overestimate the current
level, as well as in the weak coupling regime, where the
observed energy gaps are much smaller than expected.
Experiments on organic molecules have shown that the so-called
addition energy, which is the difference between the molecular
ionization potential (IP) and the electron affinity (EA), is heavily
reduced compared to its gas phase value in single-molecule
SETs~\cite{Bjornholm-OPV5,vanderZant-Vibration,Bjornholm-OPV3}.

Reductions of the excitation gaps is well-known from other situations.
Theoretical studies of semiconductor/metal interfaces have shown that the band-gap
of the semiconductor is narrowed near the interface by the screening in the
metal~\cite{Inkson,Godby-AlGaAs}. Experiments using photoemission techniques and
scanning tunneling spectroscopy to study the electronic structure of
single molecules, self-assembled monolayers and organic thin
films on dielectric and metal surfaces have shown similar
effects~\cite{Hesper-C60,Kahn-ThinFilms,Jonkman-OPV5C60,
Zhu-SpectroscopicView,Louie-C60OnMetal,Repp-PentaceneOnNaCl}.
The experimental settings of a single-molecule
SET is to some degree analogous to the situation in electrochemical measurements
where the equivalent of the addition energy, the electrochemical gap, is
well-known to depend on the dielectric properties of the surrounding media~\cite{Marcus}.
However, the screening environments are rather different for
the electrochemical setup and the single-molecule transistor
geometry, with one being in ionic solutions or organic solvents
and the other in solid state low temperature environment.
A direct comparison is therefore not possible in general.

It has been suggested that the reduction of the addition energy
seen in single-molecule SETs is caused by polarization/image charges in the
metallic electrodes~\cite{Bjornholm-OPV5,Hedegaard-OPV5},
giving rise to a localization of the charges near the metallic
electrodes. Theoretically only a few other studies have
addressed the polarization induced renormalization of the molecular levels
in solid state environments and its implications for the electron
transport in molecular junctions~\cite{Louie-Benzene,Ratner-Polarization,
Neaton-AmineGold,Thygesen-FunctionalGroups}, and the situation is still very much debated.
Therefore, a more realistic and quantitative theoretical description of the
surprisingly large effect is called for. The purpose of the
present letter is to fill out this gap and study the influence of the
junction environment on the the positions of the molecular
levels in a realistic single-molecule SET. We have developed
a scheme that includes the polarizable environment in a quantum chemical
calculation, in which the static polarization response of the environment and
the molecular charge distribution is determined self-consistently.
Our calculations on the conjugated organic molecule used in
experiments~\cite{Bjornholm-OPV5,vanderZant-Vibration} show that a large part of
the reduction of the addition energy can be accounted for by polarization of
the environment. By using a simplified expression for the addition energy, the
reduced addition energy can be understood in terms screening of the charging
energy of the molecule and a closing of the HOMO-LUMO gap.
\begin{figure*}
  \centering
  \includegraphics[scale=1.]{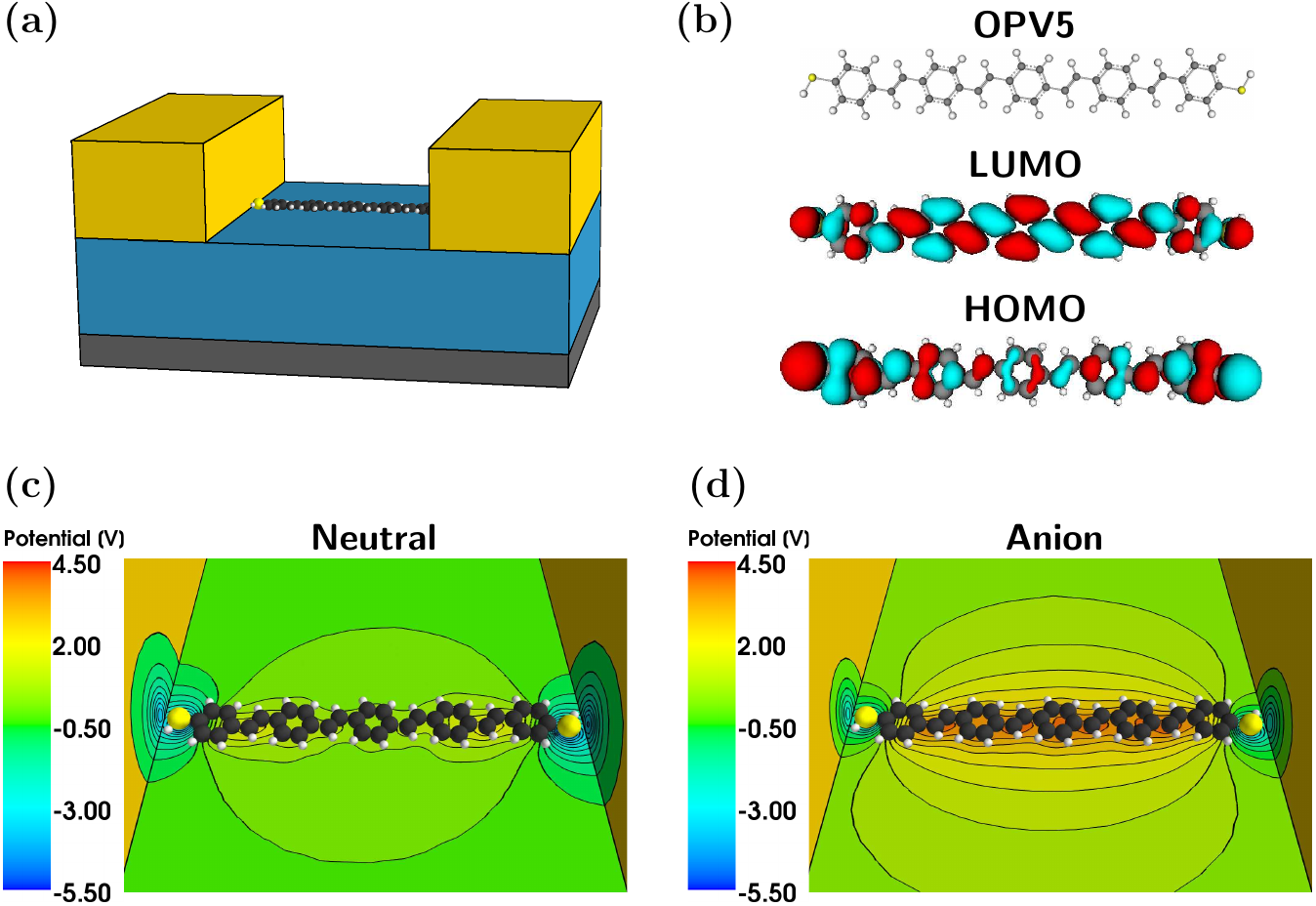}
  \caption{\label{fig:fig2}
    (a) Illustration of the simulated OPV5-SET with the molecule lying
    flat on the Al$_2$O$_3$ gate oxide between the source and drain electrodes.
    (b) Molecular structure of OPV5 and isosurface plots (blue: negative, red: positive)
    of the HOMO and LUMO orbitals for the isolated molecule.
    (c),(d) Polarization response of the nanojunction illustrated by contour plots of
    the induced potential from the polarization charge
    at the dielectric and electrode interfaces for the neutral molecule (c) and for the anion (d).
    Due to its partially positive charged thiol groups the overall
    neutral OPV5 molecule induces a negative potential in the nearby electrodes
    and dielectric that reduces the HOMO-LUMO gap (see text).
    The additional electron of the anion results in a
    significant polarization of the gate dielectric which reduces the
    charging energy of the molecule.}
\end{figure*}

For a single-molecule SET operating in the Coulomb blockade regime, i.e. with
a weak tunnel coupling between the molecule and the source/drain electrodes,
sequential tunneling is the dominating transport mechanism. In this regime
charge transfer through the molecule is possible when either the IP or the EA
is positioned within the bias window. If on the other hand no levels are present
in the bias window, current is blocked and the molecule remain in a fixed
charged state.
To reach the regime where tranport is possible one can either shift the IP and EA
levels with the gate-voltage or apply a sufficiently large source-drain bias.
This results in a so-called charge stability diagram which maps out the molecular
charged states as a function of source-drain and gate voltage (see e.g.
Figure~\ref{fig:fig4}). The addition energy $U$ can be extracted from this diagram
by measuring the height of the central diamond. Since the ionization potential
and the electron affinity are given by the difference in total energy between
the neutral molecule (with $N$ electrons) and the cation ($N-1$) and anion ($N+1$),
respectively,
\begin{equation}
  \label{eq:eq1}
  \mathrm{IP} = E^{N-1} - E^N \quad  \mathrm{and} \quad
  \mathrm{EA} = E^N - E^{N+1}
\end{equation}
the addition energy can be expressed as,
\begin{equation}
  \label{eq:eq2}
  U = \mathrm{IP} - \mathrm{EA} = E^{N+1} + E^{N-1} - 2E^N.
\end{equation}
When the molecule is placed in a nanojunction, charging of the molecule induces
polarization charge in the junction environment. The formation of the polarization charge
is associated with stabilizing polarization energies $P_+$ (added hole) and $P_-$
(added electron) for the cation and anion, respectively~\cite{Kahn-ThinFilms}, which
shifts the IP and EA relative to their gas phase values as illustrated in Figure~\ref{fig:fig1}b.
The resulting reduction of the addition energy is given by the sum $P = P_+ + P_-$,
i.e. $U = \mathrm{IP(g)} - \mathrm{EA(g)} - P$. Naturally, the polarization energy
$P$ depends on the screening properties and response times of the environment.
In single-molecule SETs where the typical current level is on the order of
$I\sim \mathrm{pA-nA}$,
the polarization response of the metallic electrodes and gate dielectric (given
by the plasmon frequency $\sim 10^{15}\;\mathrm{s}^{-1}$ and the phonon frequency
$\sim 10^{13}\;\mathrm{s}^{-1}$ respectively) is orders of
magnitudes faster than the tunneling rate $\Gamma = I/e\sim 10^{7}-10^{10}\;\mathrm{s}^{-1}$
for electrons, implying that the polarization energy is given by
the full static response of the environemnt.

In the Appendix we have provided a general framework for
evaluating total energies of nanoscale systems in the presence
of a polarizable environment. The main assumption of our
approach is that the polarizable environment responds
instantaneously to changes in the charge state of the molecule,
which according to the above consideration is a reasonable
assumption for single-molecule SETs. An electrostatic treatment
of the environment hence suffices and we derive the following
effective Hamiltonian for the nanojunction,
\begin{equation}
  \label{eq:eq3}
  H = H_{_S} + H_{pol} + H_{ext}
\end{equation}
where $H_{_S}$ is the Hamiltonian of a general nanoscale system (in our
case a molecule),
\begin{equation}
  \label{eq:eq4}
  H_{pol} = \int \! d\mathbf{r}\; \rho_{_S}(\mathbf{r})\Phi_{ind}(\mathbf{r})
            - \frac{1}{2} \int \! d\mathbf{r}\;\langle \rho_{_S}(\mathbf{r}) \rangle
            \Phi_{ind}(\mathbf{r})
\end{equation}
describes the interaction between the molecular charge distribution $\rho_{_S}$
and the polarization charge through the induced potential $\Phi_{ind}$, and
\begin{equation}
  \label{eq:eq5}
  H_{ext} = \int \! d\mathbf{r}\; \rho_{_S}(\mathbf{r})\Phi_{ext}(\mathbf{r})
\end{equation}
accounts for external voltages applied to the gate, source and drain electrodes.
The external potential $\Phi_{ext}$ satifies Laplace equation with boundary-conditions
given by the applied voltages on the electrodes.
The induced potential $\Phi_{ind}$ can be obtained via a solution to Poisson's equation
\begin{equation}
  \label{eq:eq6}
  -\nabla \cdot \left[ \varepsilon_r(\mathbf{r}) \nabla \Phi
    (\mathbf{r})\right] =  4\pi \rho_{_S}(\mathbf{r}) ,
\end{equation}
for the potential $\Phi = \Phi_{_S} + \Phi_{ind}$, where $\varepsilon_r$ is the dielectric
constant of the enviroment and $\Phi_{_S}$ the potential from the molecular charge
distribution.
The present approach thus allows for a continuum description of
the environment combined with a quantum chemical treatment (e.g. DFT or Hartree-Fock)
of the molecule. In order to account for the molecular charge redistribution
due to the polarization response of the environment the induced potential must
be included in the usual self-consistent cycle of e.g. DFT calculations.

In the present work a semi-empirical method has been combined
with a finite element treatment of Poisson's equation (see
Appendix A for details). We note that the addition energy we
calculate for the isolated OPV5 molecule (see below), is
underestimated with $1.2\;\mathrm{eV}$ as compared to the DFT
value using the B3LYP exchange-correlation functional. In spite
of this fact, we still expect the polarization energies to be
accurate, since the interaction with the polarization charge is
treated correctly in our approach. That this is indeed the case
has been confirmed by comparison with other
methods~\cite{Louie-Benzene}.

We apply here our method to a single-molecule SETs based on the thiol-terminated
OPV5 molecule, which is a organic conjugated molecule consisting of alternating
phenylene and vinylene groups (see Figure~\ref{fig:fig2}b).
In experimental realizations of OPV5-based SETs both heavily reduced
addition energies, acces to several redox states, molecular vibrational excitations
and Kondo effect have been observed~\cite{Bjornholm-OPV5,vanderZant-Vibration,
vanderZantPaaske}.
The simulated OPV5-SET is illustrated in Figure~\ref{fig:fig2}a with
the molecule lying flat on the gate dielectric between the source and
drain electrodes, which are separated by a $3.2\;\mathrm{nm}$ gap.
The gate electrode is separated from the molecule by a $5\;\mathrm{nm}$ thick layer
of gate oxide with dielectric constant $\varepsilon_r = 10$, corresponding
to the high-$\kappa$ dielectric Al$_2$O$_3$ often used in experiment. The relatively
high dielectric constant of Al$_2$O$_3$ ensures a reasonable capacitive coupling to the gate
electrode. The gold electrodes are modelled by infinitely high metal blocks.
The molecule is placed at a distance of $1\;\mathrm{\AA}$ from the surfaces of the
source/drain electrode and the gate oxide. Since the electrostatic image plane of
atomic surfaces is located outside the atomic surface
plane~\cite{LamNeeds,Chulkov-image}, this effectively corresponds to
a distance between the molecule and the surface atoms on the order of van der
Waals distance ($\sim 3\;\mathrm{\AA}$).

Table~\ref{tab:tab1} summarizes our findings for the addition energy,
single-particle HOMO-LUMO gap and polarization energy in the following three
environments: (i) gas phase, (ii) as in Figure~\ref{fig:fig2}a and (iii) molecule placed
in the gap between two parallel metal surfaces separated by $3.2\;\mathrm{nm}$.
The polarization energies due to the presence of the
junction environments results in significant reductions of the addition energies
relative to their gas phase values.

For further analysis we shall use the following simplified interpretation of
the addition energy: starting with two neutral molecules then $U$ is the energy
cost of transferring an electron from one molecule to other
(see equation~\eqref{eq:eq2}).
Since this process involves the promotion of an electron from the HOMO
in one of the molecules to the LUMO in the other molecule, it is suggestive
to write the addition energy as the HOMO-LUMO gap of the neutral molecule,
$\Delta_{HL}$, plus two times the Coulomb energy, $E_c$, required
to charge a molecule
\begin{equation}
  \label{eq:eq7}
  U = \Delta_{HL} + 2 E_c .
\end{equation}
This is similiar to the expression for the addition energy in the
constant-interaction model, which has been used successfully for
conventional quantum dot SETs~\cite{Kouwenhoven}.

In a naive first guess one would expect the reduction of $U$ to be mainly a consequence
of screening of the charging energy $E_c$.
However, Table~\ref{tab:tab1} shows that also the HOMO-LUMO gaps are reduced in the polarizable
environments. The origin of this reduction is
illustrated in Figure~\ref{fig:fig2}c. Due to the positively charged thiol groups of the
overall neutral OPV5, a negative electrostatic potential is induced in the nearby
electrodes and dielectric. This combined with the localization of the HOMO
on the thiol groups, see Figure~\ref{fig:fig2}b, shifts the HOMO level to higher energy.
Similar reasoning for the negatively charged carbon backbone and the LUMO
leads to a lowering of the LUMO level and hence a closing of
the HOMO-LUMO gap. The charging energy obtained from equation~\eqref{eq:eq7}
is for OPV5 in gas phase $E_c=1.08\;\rm{eV}$. The screening response of the
nanojunction, which is shown for the OPV5 anion in Figure~\ref{fig:fig2}d, reduces
this value to $E_c=75\;\rm{meV}$. We can therefore understand the reduction of
the addition energy as a consequence of two parallel
effects: i) a reduction of the HOMO-LUMO gap and ii) screening of the
Coulomb repulsion on the molecule which lowers the charging energy.
Since the majority of the reduction is due to the latter effect which is
purely electrostatic in nature, the reduction for other molecules of
the same size of OPV5 will be comparable. For smaller molecules we have
found that the closer proximity of the polarization charge enhances the
screening of the charging energy, resulting in a larger absolute reduction.
\begin{table}
\begin{center}
\renewcommand{\arraystretch}{1.4}
\begin{tabular}{|c||ccc|} \hline 
environment   &   U  & $\Delta_{HL}$ &  $P$  \\ \hline \hline
gas phase     & 3.27 &    1.12       &   -   \\
SET           & 0.68 &    0.53       &  2.59 \\
gap           & 2.08 &    0.92       &  1.19 \\ \hline
\end{tabular}
\caption{\label{tab:tab1}
  Calculated addition energies, $U$, single-particle HOMO-LUMO gaps, $\Delta_{HL}$,
  and polarization energies, $P$, (all in $\rm{eV}$) for the
  thiol-terminated OPV5-molecule in the three geometries: gas phase
  (isolated molecule), SET (geometry as in Figure 2a),
  and gap (molecule placed in the gap between two infinite parallel metal surfaces).}
\end{center}
\end{table}
\begin{figure*}
  \centering
  \includegraphics[scale=1.]{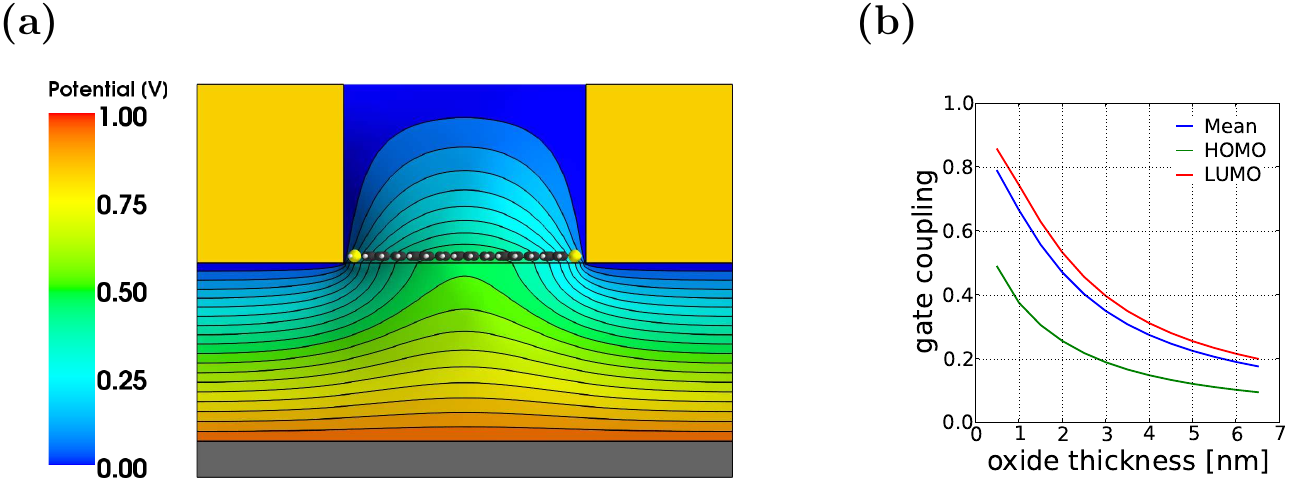}
  \caption{\label{fig:fig3}
    (a) Contour plot of the gate potential for an oxide thickness of $2.5\;\rm{nm}$
    and with $1\;\rm{V}$ applied to the gate electrode. The metallic electrodes,
    which are held at $0\;\rm{V}$, screen the gate potential
    significantly which results in a stronger coupling to the central part of the
    molecule. (b) Gate couplings as a function of oxide thickness. The gate couplings
    have been calculated as follows: (Mean) by averaging the potential over the atomic
    positions of the molecule; (HOMO) and (LUMO) by calculating how much the levels move;
    both with $1\;\rm{V}$ applied to the gate electrode.
    The localization of the HOMO on the thiol groups results in a much lower gate
    coupling compared to the LUMO, which is delocalized over the carbon backbone of
    the molecule. For an oxide thickness of $5\;\rm{nm}$ the gate coupling (Mean)
    of $\sim 0.2$ is in good agreement with the value reported in the
    experiment~\cite{Bjornholm-OPV5}.}
\end{figure*}

The important role of the gate oxide in the reduction of $U$ is
clearly demonstrated by the large difference in the polarization energy between
the SET and gap environment in Table~\ref{tab:tab1}. More than half of the polarization
energy of $2.59\;\mathrm{eV}$ in the SET environment is due to the gate oxide.
As the molecule is lying flat on the oxide which has almost metallic-like screening properties
(the image charge strength of a dielectric surface is
$q_{\varepsilon_r} = \varepsilon_r/(\varepsilon_r+1)$)
this should come as no surprise. Note, however, that the polarization energy in the SET
environment is highly dependent on the dielectric constant of the gate oxide. With a
SiO$_2$ oxide layer ($\varepsilon_r=3.9$) the polarization energy
is $2.11\;\rm{eV}$. Due to its large distance to the molecule the
gate electrode has no effect on the polarization energy.

As mentioned in the introductory part, the molecular levels can
be probed by shifting them with the gate voltage or opening the
source-drain bias window. Single-molecule SETs therefore provide
a usefull tool for measuring the energy differences between the
molecular levels through the charge stability diagram, albeit in
an unnatural environment. We have calculated the charge stability
diagram for the OPV5 device by evaluating total energies of the
neutral, singly charged and doubly charged molecule as a function
of gate and source-drain voltage. The resulting stability diagram
shown in Figure~\ref{fig:fig4} is in qualitative agreement with
experimental results~\cite{Bjornholm-OPV5}.
It is characterized by two small diamonds enclosing a big central diamond.
The hight of the central diamond is seen to be $\sim 0.50\;\mathrm{eV}$
instead of $0.68\;\mathrm{eV}$ as we found for the zero bias value of
the addition energy in Table~\ref{tab:tab1}. This is due to the fact
that the HOMO level moves downwards with the applied source-drain voltage,
and hence decreasing the threshold for pulling out an electron from the HOMO.
The non-linear edges on the left side of the central diamond is a result
of this effect.
The heights of the two smaller diamonds correspond to the addition energies
of the anion and cation of the OPV5 molecule, i.e.
\begin{equation}
  \label{eq:eq8}
  U^{N\pm 1} = E^{N}+E^{N\pm 2}-2E^{N\pm 1} .
\end{equation}
The small addition energies associated with these states stem from their
half filled frontier orbitals. Therefore, when adding/removing an electron
to/from the anion/cation only the charging energy in equation~\eqref{eq:eq7}
contributes.
The resulting charging energies are $\sim 50\;\rm{meV}$ and
$\sim 85\;\rm{meV}$, respectively, showing that due to the different spatial
distributions of the HOMO and LUMO, the charging energy of the cation and anion
are not equal, which is implicitely assumed in equation~\eqref{eq:eq7}

One important ingredient in understanding the stability diagram is
the gate coupling, $\alpha = \partial E_{mol}/\partial V_g$, i.e. how much the
energy landscape on the molecule changes when a voltage is applied to the gate
electrode. For usual quantum dot devices this is characterized
by a single number, which assumes that all states couple equally to the gate.
For the OPV5-SET considered here this is not the case. As shown in
Figure~\ref{fig:fig3}a, the gate potential varies significantly over the
extend of the molecule due to screening in the metallic electrodes, which results
in a higher gate coupling to the LUMO compared to the HOMO.
In the stability diagram this is reflected in the different slopes of
the diamond edges, which are given by the gate couplings to the different
charged states of the molecule. This has also been observed in a recent
experiment~\cite{vanderZantPaaske}. The slopes of the diamond edges agree well
with the calculated gate couplings in Figure~\ref{fig:fig3}b, where we read
off the values $\alpha_{_{\mathrm{HOMO}}} \sim 0.12$ and
$\alpha_{_{\mathrm{LUMO}}} \sim 0.25$ for an oxide thickness of
$5\;\mathrm{nm}$.
\begin{figure}
  \centering
  \includegraphics[scale=1.]{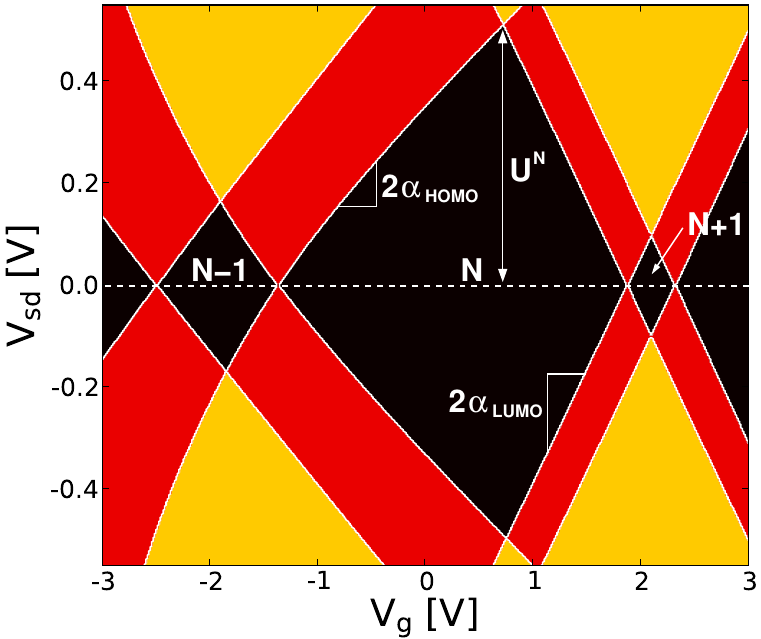}
  \caption{\label{fig:fig4}
    Charge stability diagram for the OPV5-SET. The color indicates
    the number of levels positioned in the bias window (black: 0, red: 1,
    yellow: 2) and is hence an indirect measure of the current level for
    a given gate and source-drain voltage. The Fermi levels of the gold
    electrodes have been placed in the gap of the molecule as illustrated
    in Figure~\ref{fig:fig1}b.
    In each of the black diamonds current is blocked leaving the molecule
    in the indicated charge states. The addition energies of the charge states
    of the molecule can be read off from the heights of the respective diamonds.}
\end{figure}

In conclusion, by using a method where a continuum description
of the polarizable junction environment is combined with a quantum chemical
calculation for the molecule, we have studied the effect of
polarization in an OPV5 single-molecule SET.
Our results show a significant modification of the addition energy
caused by both a closing of the HOMO-LUMO gap and screening of the
intra-molecular Coulomb interactions.
Since the majority of the reduction is due to the latter effect,
the reduction obtained here for OPV5, should be general for other
molecules of same size.
From the calculated charge stability diagram we explain at a
qualitative level the origin of alternating diamond sizes,
state dependent gate couplings and non-linear diamond
edges, which all have been observed experimentally.

Our calculations explain a large part of the reductions observed
exerimentally, but certainly not all. Other effects not accounted
for in the present work that can reduce the addtion energy even
further could be i) a geometry where the molecule is closer the
the metallic electrodes as compared to our idealized setup ii)
polaron formation upon charging of the molecule, i.e. relaxation of
the nuclear configuration, which is associated with relaxation
energies on the order of $\sim 200-300\;\mathrm{meV}$~\cite{Geskin-PolaronOPV}
iii) a correlation induced localization of the added charge
near the metallic electrodes beyond what can be captured by
a mean-field approach.

{\bf Acknowledgement.} We would like to thank T.~Bj{\o}rnholm for illuminating discussions on
the chemical aspects of the problem, K.~Stokbro for useful inputs on the
numerical calculations, Atomistix A/S for providing numerical
routines for electronic structure calcalculations and people involved in the
FEniCS project for providing useful insight in their FEM software.
Financial support from the Danish Council for Production and
Technology (FTP) under grant 26-04-0181 `Atomic scale modeling of
emerging electronic devices' and the European Community's Seventh Framework
Programme (FP7/2007-2013) under grant `SINGLE' no 213609 is acknowledged.


\appendix
\section{Supporting Information}

\subsection{Theoretical Framework} Here we show how to include
the polarizable environment of a molecular single-electron
transistor (SET) in a total energy calculation of the molecule.
The main idea is the following: when treating the metallic
leads and gate dielectric in a continuum description they enter
only in the molecular Hamiltonian as an effective potential.

In order to address the position of the ionization potential
and electron affinity in the molecular junction, the full
junction Hamiltonian (i.e. molecule, metallic leads, gate
dielectric and their mutual interaction) must be considered
\begin{equation}
  \label{eq:H_tot}
  H = H_{_S} + H_{_E} + H_{_{SE}}.
\end{equation}
The first term is the Hamiltonian of a nanoscale system $S$
(here a molecule). In general, $H_{_S}$ is a quantum mechanical
many-body Hamiltonian that can be treated with quantum chemical
methods of required accuracy. For now $H_{_S}$ will not be
specified further. The second term is the Hamiltonian of the
polarizable environment ($E$), which accounts for the energy
cost due to build-up of the polarization charge in the metallic
electrodes and the gate dielectric. Since the dynamical
polarization response of the environment is orders of magnitude
faster than the electronic tunneling rates between the molecule
and leads, an electrostatic treatment of the interaction will
be sufficient. Therefore we replace the environments by their
classical electrostatic energies, which are functionals of the
charge distributions and polarizations in the leads and in the
dielectric. This approximation neglects the kinetic energy and
the exchange-correlation energy associated with the charge
build up.

For the metallic environment where the electrons are free to
move around the electrostatic energy is
\begin{equation}
  \label{eq:H_m}
  H_{m} = \frac{1}{2} \int\! d\mathbf{r} \! \int\! d\mathbf{r}'\;
           \rho_{m}(\mathbf{r}) V_{_C}(\mathbf{r}-\mathbf{r}')
           \rho_{m}(\mathbf{r}'),
\end{equation}
where $\rho_m$ is the charge density of the metal and
$V_{_C}(\mathbf{r}) = 1/\vert\mathbf{r}-\mathbf{r}'\vert$ is
the Coulomb interaction (atomic units are used throughout this
note if not otherwise stated). In the dielectric environment
the polarization charge is bound in small dipoles
$\mathbf{p}_i$, giving rise to a macroscopic polarization
$\mathbf{P}=\sum_i \mathbf{p}_i$ of the dielectric. The energy
of a polarized dielectric can be expressed as
\begin{eqnarray}
\nonumber
  H_{d} &=& \frac{1}{2} \int\! d\mathbf{r} \! \int\! d\mathbf{r}'\;
           \rho_{b}(\mathbf{r}) V_{_C}(\mathbf{r}-\mathbf{r}')
           \rho_{b}(\mathbf{r}')\\&& +
           \frac{1}{2} \int\! d\mathbf{r} \frac{4\pi}{\chi
             (\mathbf{r},\mathbf{r})}
           \vert \mathbf{P}(\mathbf{r}) \vert^2 ,  \label{eq:H_d}
\end{eqnarray}
where $\rho_b=-\nabla \cdot \mathbf{P}$ is the bound charge of
the dielectric and $\chi$ the electric susceptibility. The
first term in equation~\eqref{eq:H_d} describes the
electrostatic dipole-dipole interaction. The second local term,
which is equivalent to the energy of a spring
$\frac{1}{2}kx^2$, accounts for the energy stored in the
dipoles $\mathbf{p}_i$. Utilizing that $\nabla \cdot
\frac{\mathbf{r-r}'}{\vert \mathbf{r-r}'
\vert^2}=\delta(\mathbf{r-r}')$ this term can recast in the
form $ \frac{1}{2} \int\! d\mathbf{r} \! \int\! d\mathbf{r}'\;
\rho_{b}(\mathbf{r}) \tilde{V}(\mathbf{r};\mathbf{r}')
\rho_{b}(\mathbf{r}') $, where $\tilde{V}$ is a complicated
interaction between the bound charges. The total Hamiltonian of
the environment can now be written
\begin{equation}
  \label{eq:H_E}
  H_{_E} = \frac{1}{2} \int\! d\mathbf{r} \! \int\! d\mathbf{r}'\;
          \rho_{_E}(\mathbf{r}) V(\mathbf{r};\mathbf{r}')
           \rho_{_E}(\mathbf{r}'),
\end{equation}
where $\rho_{_E} = \rho_{m} + \rho_{d}$ and $V$ is either
$V_{_C}$ or $V_{_C}+\tilde{V}$. As we shall see in the
following, a further specification of $V$ is not required. The
last term in equation~\eqref{eq:H_tot} accounts for the
interaction between the molecule and the environment. Since we
are focusing on the Coulomb blockade regime where the molecule
has small tunnel couplings to the leads, the hybridization term
of $H_{_{SE}}$ can be neglected. What remains is the Coulomb
interaction between the spatially separated charges of the
molecule and of the environment
\begin{equation}
  \label{eq:H_SE}
  H_{_{SE}} = \int\! d\mathbf{r} \! \int\! d\mathbf{r}'\;
              \rho_{_S}(\mathbf{r}) V_{_C}(\mathbf{r}-\mathbf{r}')
              \rho_{_E}(\mathbf{r}').
\end{equation}
Notice that the molecular charge has contributions from both
the ionic cores and the valence electrons of the molecule:
$\rho_{_S}(\mathbf{r})= \rho_{ion}(\mathbf{r}) +
\rho_{e}(\mathbf{r}) =\sum_i Z_i
\delta(\mathbf{r}-\mathbf{r}_i)-\psi^{\dagger}(\mathbf{r})\psi(\mathbf{r})$.

With the Hamiltonian in place we now proceed to eliminate the
environment degrees of freedom. Since the part of the
Hamiltonian involving $\rho_{_E}$ is classical and has no
dynamics, the solution for $\rho_{_E}$ can be found by
minimizing with respect $\rho_{_E}$, i.e. by setting $\delta
H/\delta\rho_{_E}=0$. The resulting equation can be solved for
$\rho_{_E}$ giving (in matrix notation)
\begin{equation}
  \label{eq:InducedCharge}
  \rho_{_E} = - [V_{_{EE}}]^{-1} V_{_{ES}} \rho_{_S} \equiv \rho_{ind},
\end{equation}
which represents the charge density induced by the system
charge $\rho_{_S}$. Inserting this expression for the induced
charge of the environment back into the full Hamiltonian in
equation~\eqref{eq:H_tot}, the terms involving $\rho_{_E}$ can
be recast in the form
\begin{align}
  \label{eq:EffectiveInteraction}
  H_{_E} + H_{_{SE}} = -\frac{1}{2} \rho_{_S} V_{_{SE}}
                         \left(V_{_{EE}}\right)^{-1} V_{_{ES}}\rho_{_S},
\end{align}
in which the degrees of freedoms of the environment has been
substituted by an effective interaction between molecular
charges. By inserting the expression in
equation~\eqref{eq:InducedCharge} for the induced charge
density into equation~\eqref{eq:EffectiveInteraction}, we find
that $H_{_E} + H_{_{SE}} = \frac{1}{2} \rho_{_S} V_{_{SE}}
\rho_{ind}$, which is one half of $H_{_{SE}}$ in
equation~\eqref{eq:H_SE}. This result, which is a
generalization of the classical image charge problem (a point
charge placed at a distance $z$ from a perfect conducting
surface where $H_{_E} + H_{_{SE}}= -1/4z$), states that the
energy cost associated with the build up of the polarization
charge is always one half of the energy gained by the system
$S$ through its interaction with the polarization charges.

Due to the presence of the electronic field operators in
$\rho_{_S}$ the effective interaction in
equation~\eqref{eq:EffectiveInteraction} must be approximated.
In our scheme we use a Hartree approximation. This is justified
because i) correlation effects due to the effective interaction
are small given the absence of a short range interaction and
ii) exchange is not relevant since self-interactions are
possible via the image charges. Introducing the induced
potential $\Phi_{ind}(\mathbf{r}) =\int\! d\mathbf{r}'
\rho_{ind}(\mathbf{r}') /\vert \mathbf{r} - \mathbf{r}' \vert$,
the Hartree version of equation~\eqref{eq:EffectiveInteraction}
can be written as
\begin{equation}
  \label{eq:EffectiveHamiltonian}
  H_{_E} + H_{_{SE}} = \int \! d\mathbf{r}\; \rho_{_S}(\mathbf{r})
                       \Phi_{ind}(\mathbf{r})
                       - \frac{1}{2} \int \! d\mathbf{r}\;
                       \langle \rho_{_S}(\mathbf{r}) \rangle \Phi_{ind}(\mathbf{r}),
\end{equation}
where the last term subtracts the double counted contributions
to the total energy in the first term.

The induced potential can be obtained by solving Poisson's
equation
\begin{equation}
  \label{eq:PoissonsEquation}
  -\nabla \cdot \left[ \varepsilon_r(\mathbf{r}) \nabla \Phi_{tot}
    (\mathbf{r})\right] =  4\pi \rho_{_S}(\mathbf{r})
\end{equation}
with Dirichlet boundary-conditions on the electrode surfaces
$S_i$, i.e. $\Phi_{tot}=0$ if $\mathbf{r}\in S_i$. Here, the
total potential is the sum of the potential from the system
charges plus the potential from the induced charges:
$\Phi_{tot} = \Phi_{_S} + \Phi_{ind}$.

The present approach thus allows us to combine a continuum
description of the junction environment with a quantum chemical
description of the molecule.

\subsection{Semi-empirical Method} The molecular part of the
total junction Hamiltonian in equation~\eqref{eq:H_tot} is
split up into two parts: i) one that accounts for the isolated
molecule and ii) one that accounts for
polarization/redistribution of charge due to interactions with
the junction environment and/or charging of the molecule,
\begin{equation}
  H_{_S} = H_0 + H_{pol}.
\end{equation}
In the following our semi-empirical treatment is descriped in
detail. It is similiar to the one presented in
Ref.~\cite{Zahid}, however, here applied to total energy
calculations and generalized to include dielectrics in the
determination of the induced potential $\Phi_{ind}$.

The part describing the isolated molecule is an effective
tight-binding Hamiltonian
\begin{equation}
  \label{eq:H_isolated}
  H_0 = \sum_i \varepsilon_i c^{\dagger}_i c_i + \sum_{i\ne j}t_{ij} c^{\dagger}_i c_j ,
\end{equation}
where the sums run over atomic valence orbitals $\{\phi_i\}$
and $i$ is a collective index referring to atom, orbital and
spin index: $i \rightarrow \mu i \sigma$. We use Extended
H{\"u}ckel parameters by Hoffmann~\cite{Hoffmann} for the
onsite and hopping energies
\begin{eqnarray}
   \varepsilon_i & = & -V_i \\
   t_{ij} & = & \frac{1}{2} k S_{ij} \left( \varepsilon_i + \varepsilon_j \right) .
\end{eqnarray}
Here $V_i$ is associated with the ionization energy of the
valence orbital $\phi_i$, $k$ is a fitting parameter usually
set to $1.75$ and $S_{ij} = \langle \phi_i \vert \phi_j
\rangle$ is the overlap between the non-orthogonal atomic
orbitals. Notice that electron-electron interactions are
implicitly included in $H_0$ due to its parametrized form.

In the part of the Hamiltonian that accounts for polarization
and charging of the molecule, electron-electrons interaction
are treated at the Hartree level. Since the Hartree potential
of the isolated molecule is indirectly accounted for in $H_0$,
only changes in the Hartree potential due to variations in the
electron density from its value, $n_0$, in the isolated
molecule are considered
\begin{equation}
  \label{eq:V_H}
  \delta V_{\text{H}}(\mathbf{r}) = \int d\mathbf{r}' \;
         \frac{\delta n(\mathbf{r}')}{\vert \mathbf{r}-\mathbf{r}'\vert} ,
\end{equation}
where $\delta n = n - n_0$. Since the Hartree potential depends
on the electron density, this posses a self-consistent problem
that must be iterated to convergence.

To simplify the numerics the integral in
equation~\eqref{eq:V_H} is approximated by a sum over atomic
point charges given by the Mulliken populations
$n_{\mu}=\text{Tr}\left[\rho S\right]_{\mu}$, where $\rho$ is
the density matrix
\begin{equation}
  \label{eq:V_point}
  \delta V_{\text{H}}(\mathbf{r}) = \sum_{\mu}
       \frac{\delta n_{\mu}}{\vert \mathbf{r}-\mathbf{r}_{\mu}\vert} .
\end{equation}
To avoid problems with the diverging point charge potential
when evaluated at the atomic positions, $\mathbf{r}_{\mu}$, the
onsite contribution to the sum is replaced by a species
dependent Hubbard U taking into account the energy cost of
adding an electron to the atom. These parameters are taken from
the quantum chemical CNDO method~\cite{Fulde,PopleSegal}. In
order to keep consistency between the onsite and the offsite
interactions, the Magata-Nishimoto~\cite{SichelWhitehead}
interpolation formula is used for the latter
\begin{equation}
  U_{\mu\nu} = \frac{1}{R_{\mu\nu}+\frac{2}{U_{\mu}+U_{\nu}}} .
\end{equation}
With these approximation the final form of the Hartree
potential in equation~\eqref{eq:V_H} becomes
\begin{equation}
  \delta V_{\text{H}}(\mathbf{r}_{\mu}) = \delta n_{\mu} U_{\mu} +
                   \sum_{\nu \ne \mu} \delta n_{\nu} U_{\mu\nu}  .
\end{equation}
In our atomic basis the polarization/charging part of the
Hamiltonian is written
\begin{equation}
  H_{pol} = \sum_{i,j} V_{ij} c^{\dagger}_i c_j -
          \frac{1}{2} \sum_{\mu} n_{\mu} V_{\text{H}}(\mathbf{r}_{\mu}) ,
\end{equation}
where the last term substracts double counting in the first
term and $H_0$. The matrix representation of the Hartree
potential has been approximated as follows
\begin{align}
  \label{eq:matrixelement}
  V_i & = \langle \phi_i \vert \delta V_{\text{H}} \vert \phi_i \rangle
        \approx \delta V_{\text{H}}(\mathbf{r}_{\mu}) \nonumber\\
  V_{ij} & = \langle \phi_i \vert \delta V_{\text{H}} \vert \phi_j \rangle \approx
             \frac{1}{2} S_{ij} \left( V_i + V_j\right) .
\end{align}
The part of the Hamiltonian involving the induced potential is
written similarly
\begin{equation}
  H_{_E} + H_{_{ES}} = \sum_{i,j} V_{ij} c^{\dagger}_i c_j
      - \frac{1}{2}\sum_{\mu} \left( Z_{\mu} - n_{\mu}\right)
      \Phi_{ind}(\mathbf{r}_{\mu})
\end{equation}
with $V_{ij}=\langle \phi_i \vert V_{ind} \vert \phi_j \rangle$
evaluated as above in equation~\eqref{eq:matrixelement}.

\subsection{Poisson's Equation and $\Phi_{ind}$} In the
following section it is described how the induced potential is
determined by solving Poisson's
equation~\eqref{eq:PoissonsEquation} for the total potential
$\Phi_{tot}$.

\subsubsection{Finite element approach} One of the major
advantages of the finite element method (FEM) is its
partitioning of the solution domain into a finite number of
elements (typically triangles in 2D and tetrahedra in 3D). The
possibility to refine the element size around sharp corners and
in the vicinity of spatially rapidly varying source terms,
allows FEM to handle a large variety of problems and solution
domains of practically any geometry, hence making it very
suitable for modelling of nanoscale devices.

In the present case we seek to solve Poisson's equation in the
geometry of a single-molecule SET. To this end we use the
finite element software from the FEniCS
project~\cite{www:fenics}.

The molecular charge density, which has both electronic and
ionic contributions, is represented by a sum of atomic centered
gaussian charge distributions:
\begin{equation}
  \rho_{_S}(\mathbf{r}) = (2\pi)^{-3/2}
                          \sum_{\mu} \frac{q_{\mu}}{\sigma^3} \,
                          e^{-\vert \mathbf{r}-\mathbf{r}_{\mu} \vert ^2/2\sigma^2}
\end{equation}
Here $\sigma = a_0$ is the width of the gaussians and
$q_{\mu}=Z_{\mu}-n_{\mu}$ ($Z_{\mu}$ being the atomic valence)
is the net atomic charge.

Having obtained the total potential, $\Phi_{tot}$, the induced
potential can be extracted by substracting the potential,
$\Phi_{_S}$, from the molecular source charges, which for a
gaussian charge distribution centered at the origin is:
\begin{equation}
  \Phi_{_S}(\mathbf{r}) =
     \frac{q}{r} \text{erf}\left( \frac{r}{\sqrt{2}\sigma} \right)
\end{equation}
where $\text{erf}$ is the error function.

\begin{figure}
  \centering
  \includegraphics[width=.45\textwidth]{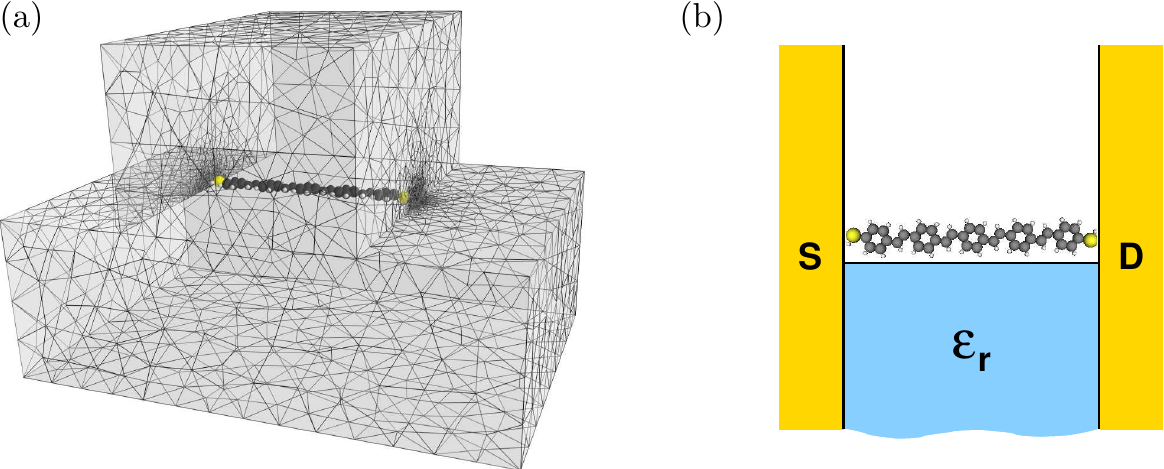}
  \caption{\textbf{Finite element mesh and simplified junction geometry.}
           (a) Surface mesh for the OPV5-SET.
           (b) Simplified junction for which Poisson's equation
           can be solved analytically. In this geometry the electrodes are
           modelled by infinite metallic surfaces and the gate oxide as a
           semi-infinite dielectric layer sandwiched between the electrode
           surfaces. The gate electrode is not included (see text).}
  \label{fig:figs1}
\end{figure}

The calculated induced potential has been converged (to
$0.05\;\rm{eV}$) with respect to the element size and the
spatial dimensions of the device. In order to get an accurate
description of the potential at the atomic positions, the
molecule is enclosed in a box with a fine mesh of element size
$\sim a_0$. The element size at the boundaries of the device
are 20 times larger, which is illustrated by the boundary mesh
in figure~\ref{fig:figs1}(a). The following device size was
used to converge the potential: an electrode hight of
$50\;\rm{a.u.}$, a device width of $100\;\rm{a.u.}$ and a
device length of $150\;\rm{a.u.}$.

\subsubsection{Analytical solution} In the simplified device
geometry shown in Fig.~\ref{fig:figs1}(b), Poisson's equation
can be solved analytically. It can be shown that the Greens
function satisfying:
\begin{equation}
  -\nabla \cdot \left[ \varepsilon_r(\mathbf{r}) \nabla
    G(\mathbf{r},\mathbf{r}')\right] = \delta(\mathbf{r}-\mathbf{r}')
\end{equation}
has the following image charge solution:
\begin{widetext}
\begin{align}
  G(\mathbf{r},\mathbf{r}')=  \sum_{\sigma=\pm 1}\sum_{\tau=\pm 1} &
        \sigma \left(\frac{\varepsilon_r+\tau}  {\varepsilon_r+1}\right)\tau
        \bigg[ \frac{1}{\sqrt{(x-\sigma x')^2 + (y-\tau y')^2 + (z- z')^2}}  \nonumber\\
      & + \sum_{n=1}^{\infty} \bigg(
          \frac{1}{\sqrt{(2nL - (x-\sigma x'))^2 + (y-\tau y')^2 + (z- z')^2}} \nonumber\\
      & + \frac{1}{\sqrt{(2nL + (x-\sigma x'))^2 + (y-\tau y')^2 + (z- z')^2}} \bigg) \bigg]
      \label{eq:analytic}
\end{align}
\end{widetext}
where $L$ is the electrode spacing ($z$ is the direction
perpendicular to the paper plane in Fig.~\ref{fig:figs1}(b))
and the sums run over all repeated images of the source charge
in the metallic surfaces and the dielectric. By leaving out the
contribution from the source charge itself, only the induced
potential remains:
\begin{equation}
  \Phi_{ind}(\mathbf{r}) = \tilde{G}(\mathbf{r},\mathbf{r}')
\end{equation}
Here $\tilde{G}$ denotes the Greens function which does not
include the $\sigma=1$ and $\tau=1$ term outside the $n$-sum in
equation~\eqref{eq:analytic}.

In the calculations using the analytical solution, the
molecular charge distribution has been approximated by point
charges $q_{\mu}$ located at the atomic positions. The induced
potential follows directly from the Greens functions of the
individual point charges:
\begin{equation}
  \Phi_{ind}(\mathbf{r}) = \sum_{\mu} q_{\mu} \tilde{G}(\mathbf{r},\mathbf{r}_{\mu})
\end{equation}

\subsubsection{FEM vs analytic} Table~\ref{tab:tabs1} summarizes
the calculated polarization energies for the realistic junction
modelled with FEM and the simplified junction shown in
Fig.~\ref{fig:figs1}(b). The difference between the
polarization energies in the two junctions is on the order of
$\sim 50\;\rm{meV}$ for the different OPV-molecules. The
slighty larger polarization energies in the simplified junction
stem from the infinite metallic electrodes, which screen the
Coulomb interactions on the molecule better than the
semi-infinite metal blocks of the realistic junction. Notice
that, due to its large distance to the molecule ($\sim
5\;\rm{nm}$) and the metallic-like screening properties of the
Al$_2$O$_3$ gate oxide, the gate electrode does not contribute
to the polarization energy in the realistic junction. This
explains the relative small differences between the
polarization energies in the two junction geometries.
\begin{table}
\renewcommand{\arraystretch}{1.4}
\begin{tabular}{||c||c|c||} \hline\hline
Molecule      &  FEM   &  Analytic \\ \hline \hline
OPV2          &  3.56  &    3.63   \\ 
OPV3          &  3.11  &    3.17   \\ 
OPV4          &  2.81  &    2.87   \\ 
OPV5          &  2.59  &    2.63   \\ \hline\hline
\end{tabular}
\caption{\label{tab:tabs1}
  Calculated polarization energies $P$ (in eV) for the two junction
  geometries illustrated in figure~\ref{fig:figs1}. The same electrode
  spacing has been used in the two geometries.
  Due to the infinite metallic surfaces, the polarization energy is slighty
  larger for the simplified junction geometry.}
\end{table}

Hence, our analytical solution provides a realistic description
of the potential in generic SET geometry (Fig.~2(a) in the main
part of the paper) that can be used instead of computationally
heavy Poisson solvers.

\providecommand{\url}[1]{\texttt{#1}}
\providecommand{\refin}[1]{\\ \textbf{Referenced in:} #1}

\end{document}